\documentclass{mem}
\usepackage{natbib}\usepackage{txfonts}\usepackage{balance}
\usepackage{graphicx}
\usepackage[a4paper]{hyperref}
\begin{document}
\def\teff{$T\rm_{eff }$}
\def\kms{$\mathrm {km s}^{-1}$}

\title{
Dynamical interstellar medium with Gaia and 
ground-based massive spectroscopic stellar surveys
}

   \subtitle{}

\author{
T. \,Zwitter\inst{1}
\and
J. \,Kos\inst{1,2}
          }

  \offprints{T. Zwitter}

\institute{
University of Ljubljana, 
Faculty of Mathematics and Physics,
Jadranska 19,
SI-1000 Ljubljana, Slovenia
\and
Sydney institute for astronomy, University of Sydney, 44-70 Rosehill Street, NSW 2016, Sydney, Australia
\\
\email{tomaz.zwitter@fmf.uni-lj.si}
}

\authorrunning{Zwitter, Kos}

\titlerunning{Dynamical ISM with Gaia and ground-based spectroscopic surveys}

\abstract{
The main observational focus of the ongoing Gaia mission of ESA is to provide an accurate spatial and kinematical information for a large fraction of stars on our side of the Galactic centre. Here we discuss how studies of general interstellar medium could reach a similar level of spatial and kinematic perfection. Interstellar extinction and line absorption studies toward a large number of stars at different distances and directions can give a 3-dimensional distribution map of interstellar absorbers. Molecular and atomic absorptions can reveal a limited dynamical picture through radial velocity measurement, but proper motion components of the velocity vector for interstellar absorbers are not measurable directly. But in the cases of special morphologies one can infer a complete velocity vector from its radial velocity component and so obtain a dynamical information comparable to stars. Such case are geometrically thin absorption curtains and sheets, where motion is likely perpendicular to the surface.

This optimistic view on what is possible builds on some prerequisites. A large number of stars at different distances need to be abserved, otherwise one cannot determine where (along the line of sight)  are the absorption pockets. This means that we should be able to measure interstellar absorptions towards cool stars which are the only ones numerous enough. It is advisable to start with interstellar absorptions at high Galactic latitudes to avoid a very complicated mix of colliding absorption clouds in the Galactic plane. But absorptions off the plane are weak, so spectra need to be well exposed. Finally, interstellar atomic line absorption studies toward cool stars in the optical are largely limited to Sodium and Potassium doublets. Many surveys, including Gaia, do not cover them. So diffuse interstellar bands (DIBs) can be important, as their measurement can give the same type of information as observation of  interstellar atomic absorption lines. A combination of DIBs and atomic absorptions may also point to differences in dynamics of different components of the interstellar medium. 

Several ongoing massive surveys obtaining medium and high-resolution stellar spectra are pursuing this approach. Gaia is a record breaker in terms of numbers of objects observed, though with a limited wavelength coverage and signal to noise. So Gaia spectra can be used to study the DIB at 8620~\AA. RAVE already demonstrated that this DIB can be used to build 3-dimensional absorption maps. APOGEE is observing a number of DIBs in the infrared, mostly along the Galactic plane. Gaia-ESO uses the Giraffe spectrograph at VLT to observe a number of DIBs in and off the plane, joined with Sodium interstellar doublet observations for its UVES observations. Finally, the Galah survey is on its way to obtain an unprecendented number of well exposed high-resolution stellar spectra off the Galactic plane which include the Potassium doublet and a dozen strong DIBs. 

Use of this new information can change our understanding in many areas. As examples we mention a new method of rejecting membership of stars in clusters, studies of a few Myr old supernova remnants and investigations of Galactic fountains.  
}
\maketitle{}

\section{Introduction}

In 2020, the Gaia mission (launched in December 2013) is expected to release astrometric distances and velocity vectors for a significant fraction of stars on our side of the Galactic centre, thus allowing a computation of stellar orbits and of evolution of the Galaxy as a whole. Studies of the interstellar medium (ISM) cannot yield information equivalent to stars, as they lack proper motion components of the velocity vector. Radial velocity shifts could be measured for $\sim 500$ diffuse interstellar bands (DIBs) identified as absorptions in the optical and infra-red spectra of background stars accumulated along the line of sight \citep{Hobbs09}. By observing a given DIB toward many stars which are nearly in the same direction but at different and known distances one can reconstruct absorption sites along the line of sight. Joining observations in many directions on the sky finally leads to their spatial distribution. So we get a 4-dimensional picture of the ISM for each DIB measurable in individual spectra. 

Interstellar absorption lines of neutral atoms yield information equivalent to DIBs, but most lines are in the UV and blue part of the spectrum, so their study is limited to hot stars. An exception are sodium and potassium doublets. These lines are sharp, so measurement of their radial velocity is easy. Absorptions by ions are mostly limited to the UV and blue domains. Interstellar emission lines lack information on the distance of the emitting clouds unless they lie in the disk and we assume they follow its rotational curve or they originate in known ISM complexes. Interstellar dust absorptions can have their spatial distribution reconstructed similarly to DIBs but they lack any velocity information, which is a prerequisite for time evolution studies. 

Here we discuss the challenge of using massive spectroscopic stellar surveys to obtain a multidimensional information on ISM, mostly based on observations of DIBs. We start with a brief overview of DIB properties and show that the ongoing surveys are reaching over a dozen DIBs in up to a million lines of sight, mostly toward stars away from the Galactic plane and with known spectrophotometric distances. ISM is the place of violent collisions of supernova shells, plus winds from asymptotic giant branch stars and hot-star associations. Head-on collisions in the Galactic plane are difficult to interpret. But many of the ongoing surveys observe away from the plane where interactions generally result in a net motion perpendicular to the plane. If any shells of absorbing material are identified we can assume that their motion is perpendicular to shell surfaces and reconstruct a complete velocity vector from its radial velocity component. Such information for ISM is then equivalent to the one collected for stars by Gaia.

\begin{table*}
\caption{Properties of selected ongoing large stellar spectroscopic surveys. 
The last column refers to spectra suitable for DIB measurements.}
\label{tablesurveys}
\begin{center}
\begin{tabular}{lccrrr}
\hline
\\
Survey & $D$(mirror) [m] & $\lambda$ range(s) [nm] & $R$ &Mag range& Stars \\
\hline
\\
RAVE    & 1.2 & 841--877 & 7500  &$9<I<12\,\,\,\,$& 459k \\
SEGUE   & 2.5 & 390--900 & 1800  &$14<g<20.3$&240k \\
SEGUE-2 & 2.5 & 380--920 & 1800  &$15.5<g<20.3$&118k \\
LAMOST  & 4.0 & 365--900 & 1000  &$r < 14\,\,\,\,$&$\gtrapprox 100$k\\
APOGEE  & 2.5 & 1510--1700&22500 &$7<H<13.8$&100k \\
Gaia-ESO& 8.0 & $\sim 80\, @$ 403--900&20000 &$V<19\,\,\,\,$&$\sim 100$k \\
GALAH   & 4.0 & $\,\, 97 \,\, @$ 472--789&28000  &$12<V<14\,\,\,$& $> 185$k\\ 
Gaia-RVS& $1.45\times 0.5$ & 847--874&11500 & $V\lesssim 13\,\,\,\,$&$\lessapprox $4M \\
\hline
\end{tabular}
\end{center}
\end{table*}


\section{Diffuse interstellar bands}

\label{secDIBs}
Diffuse interstellar bands were first discovered nearly a century ago by \citet{Heger22}  who noted the absorption bands at 5780 and 5797 \AA\ and considered them “stationary” in her study of early-type spectroscopic binaries \citet{McCall13}. They were not clearly recognised as interstellar until the work of Merrill (\citeyear{Merrill34}; see \citealt{Herbig95,Sarre06}). The label ‘diffuse’ differentiates between the somewhat hazy appearance of DIBs compared with the relative sharpness of atomic transitions in the interstellar medium. Their physical carriers are still unidentified \citep{Galazutdinov11,Krelowski10,Salama99,Snow06,Iglesias10,Maier11}. They are mostly found in the optical and near infra-red spectral bands, with the DIB with the longest wavelength discovered at 1.793 $\mu$m \citep{Geballe11}. DIBs were also observed in nearby galaxies \citep{Vidal87,Cox07,Cox14,Cordiner08a,Cordiner08b,Cordiner11} and at cosmological distances \citep[e.g.][]{York06,Monreal15}, but most of the studies rely on high resolution and high S/N spectra of hot stars in our Galaxy. Because the DIBs are weak (the strongest one at 4428 \AA\ having a typical equivalent width of 2 \AA\ in a E(B-V)=1 sight-line) and easily blended with stellar lines, high S/N spectra of nearly featureless hot stars are most appropriate for studying DIBs. Therefore most of the surveys include few thousand stars at most \citep{Snow77,vanLoon14,Lan15}, or around a hundred stars, if weaker DIBs are observed \citep[e.g.][]{Friedman11}.

\begin{figure*}[]
\resizebox{\hsize}{!}{\includegraphics[clip=true,angle=270]{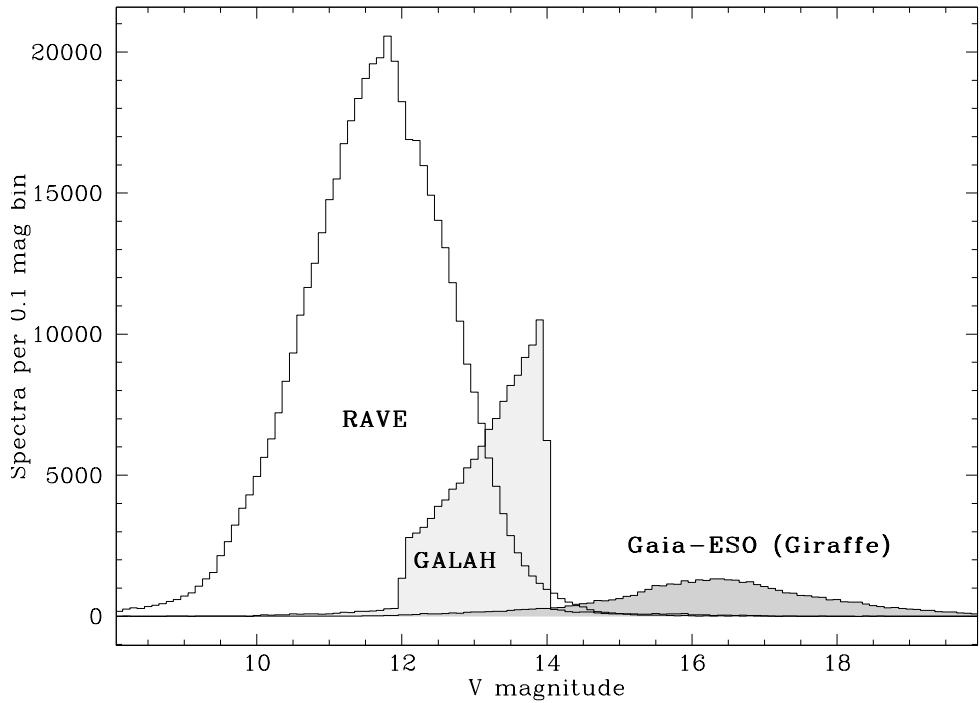}}
\caption{Apparent V magnitude distributions of RAVE, GALAH, and Gaia-ESO survey observations, the latter only for the Giraffe instrument. Only targets already observed are plotted (see text).}
\label{fig_surveys_Vhisto}
\end{figure*}

Individual DIBs do not show large variations in the peak position and profile, even between sight-lines with very different dust properties \citep[e.g. grain size, see][]{Tielens05}. Shapes of some DIBs are reminiscent of the rotational contours of a large  molecule \citep[see e.g.][]{Sarre95}.  DIB  abundances  are correlated with interstellar extinction and with abundances of some simple molecules \citep{Thorburn03}. These arguments show that DIBs are probably associated with carbon-based molecules (\citealt{Sarre06}; for a  general   review   on   carbon   role  see \citealt{Henning98}). DIBs show no polarisation effects \citep{Herbig95} and are likely positively charged \citep{Milisavljevic14}, as suggested by the relatively low energies of absorbed photons \citep{Tielens05}. Until recently no known transition of any molecule or atom has yet been found to match the central wavelengths of the DIBs 
\citep{Steglich11,Huisken14,Kokkin14,Rouille14}, a consequence of very low densities and huge absorption volumes in the ISM. Their origin and chemistry are thus unknown, a unique situation given the distinctive family of many absorption lines within a limited spectral range. Some of the DIBs have composed profiles which have the appearance of partially resolved P ($\Delta J = –1$), Q  ($\Delta J = 0$), and R  ($\Delta J = +1$) rotational branches of a large molecule, where $J$ is the molecular rotational quantum number \citep{Sarre14}. So relative positions of the components are fixed, while their relative intensities may vary. Like most molecules in the ISM that have an interlaced chemistry, DIBs may play an important role in the life-cycle of the ISM species and are the last step in fully understanding the basic components of the ISM. The problem of their identity is more intriguing given the possibility that the DIB carriers are carbon-based molecules. A recent claim of identification of DIBs at 9633~\AA\ and at 9578~\AA\ with absorption bands of C$_{60}^+$ \citep{Campbell15} may be however changing this unsatisfactory situation. 

DIBs are more numerous than absorption lines of other ISM species in the optical and near infra-red bands and are therefore ideal to be studied in general spectroscopic surveys, as they are present across the whole optical and near-IR range. Having observations of multiple DIBs also allows the study of different parameters \citep{Kos13} of the ISM apart from observing the spatial distribution of a single species. Even without the knowledge of the carriers, DIBs can be used to trace unobserved or hard to observe properties of the ISM toward the stars in a spectroscopic survey. All extensively studied DIBs correlate at least vaguely with reddening and HI abundance \citep{Herbig95,Munari08,Raimond12,Penades13}, ratios of different DIB strengths correlate with the UV radiation field \citep{Krelowski92,Kos13} and widths of some DIBs correlate with H2 abundances \citep{Gnacinski14}. It must be noted, however, that the correlation between different species of molecules can be poorly coincidential and does not indicate relations between DIBs and other species, as is often the case in the ISM.

\begin{figure*}[]
\resizebox{\hsize}{!}{\includegraphics[clip=true,angle=270]{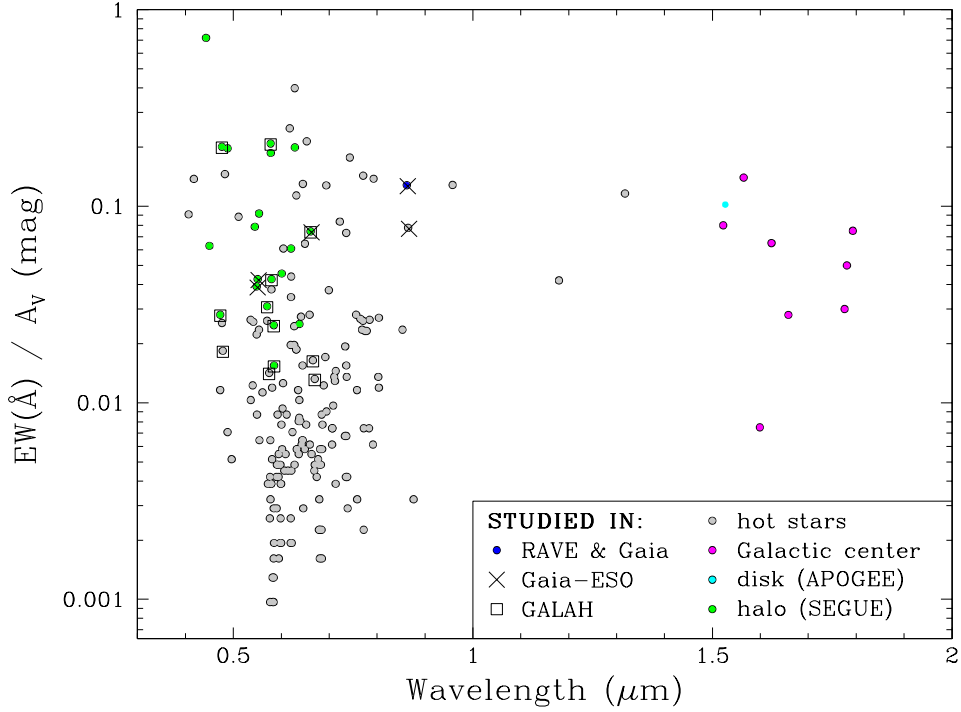}}
\caption{Distribution mean equivalent width-to-extinction (EW$_\mathrm{DIB}$/A$_V$) ratios and 
wavelength for 196 optical and NIR DIBs. Symbols mark DIBs which are being 
studied by individual spectroscopic surveys.}
\label{fig_dib_lambdas}
\end{figure*}

\section{Spectroscopic surveys}

The main goal of stellar spectroscopic surveys is to study Galactic structure and evolution. But the collected spectra allow for a significant auxiliary science
where observations of DIBs in a vast number of sight-lines are a typical example. 
Such ongoing surveys include RAVE \citep{Steinmetz06,Zwitter08,Siebert11,Kordopatis13},  SEGUE \citep{Yanny09}, SDSS-III \citep{Eisenstein11}, Gaia-ESO \citep{Gilmore12,Randich13}, APOGEE \citep{Zasowski13}, Gaia \citep{Prusti14}, Hermes-GALAH \citep{Freeman12,deSilva15} and LAMOST \citep{Deng12,Yuan14,Luo15}. Observations of 100,000s of stars bring new possibilities to the study of DIBs, to map the distribution of carriers in the Galaxy and to search for peculiar environments with unusual DIB properties. All this can contribute to the big goal of identifying the carriers. 

Table \ref{tablesurveys} lists basic properties of the ongoing large stellar spectroscopic surveys. For reasons of efficiency the exposure time per target is usually limited to $\approx 1$~hour, and most of the listed surveys use modern fiber-fed spectrographs with a high throughput. So a combination of telescope mirror diameter ($D$) and resolving power ($R$) can be used to obtain a rough estimate of the S/N ratio as a function of apparent magnitude. The number of observed stars quoted in the last column will increase with time for most of the surveys. Quoted numbers are collected from the literature for the first 5 surveys, and are current estimates for Gaia-ESo, GALAH and Gaia-RVS (counting only stars brighter than $V \sim 13$ in the latter case).

\begin{figure*}[]
\resizebox{\hsize}{!}{\includegraphics[clip=true,angle=0]{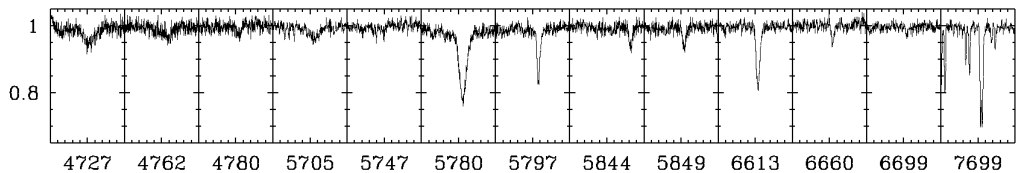}}
\caption{Preliminary data on diffuse interstellar bands and on the K I interstellar atomic line at 7699 \AA\ as observed by the GALAH survey. Each 20 \AA\ wide panel is centered on the DIB wavelength as listed in \citet{Jenniskens94}. Plotted wavelengths are heliocentric.}
\label{fig_dib_galah}
\end{figure*}

Figure \ref{fig_surveys_Vhisto} plots V magnitude histograms for stars already observed by three of the surveys listed in Table \ref{tablesurveys}. Histogram for RAVE includes stars from DR4 \citep{Kordopatis13}. Their $V$ magnitudes have been estimated from the 2MASS $J$ and $K$ magnitudes using the relation:
\begin{eqnarray}
V &=& K + 2 (J-K)+0.28 + 0.382\,  e^{2(J-K)-0.4} \nonumber \\
  & & 
\end{eqnarray}
Histogram of the GALAH survey includes stars observed until 10-apr-2015, and data for Gaia-ESO survey are the ones from iDR4 obtained with the Giraffe instrument. Another $\sim 6.5$\%\ of the Gaia-ESO targets are observed with the UVES instrument and are not plotted in Figure \ref{fig_surveys_Vhisto}. Most of the UVES targets are in the $12.5 < V < 15.5$ magnitude range. 

DIBs are distributed over a wide range of wavelengths, so surveys with a large wavelength coverage generally include a larger number of DIBs. But on the other hand most of the DIBs are very faint, so a high S/N spectrum is needed for their detection and measurement. In this sense one can note the importance of the GALAH survey which is obtaining hundreds of thousands of high resolution spectra with $S/N \sim 100$ per resolution element.

Figure \ref{fig_dib_lambdas} plots mean equivalent width to extinction ratio and wavelength of close to 200 confirmed DIBs. Grey symbols are DIBs that were studied only 
in individual spectra of hot stars. Their plotted intensities are from \citet{Jenniskens94}, \citet{Krelowski95}, and \citet{Jenniskens96}. Blue dot is the DIB at 8620~\AA\ observed by RAVE and Gaia surveys \citep{Munari99,Munari08}. Pink dots are infra-red DIBs reported in spectra of stars toward the Galactic center by \citet{Geballe11}. Light blue is the DIB at 1527~nm observed by APOGEE \citep{Zasowski15}, green dots are 
20 DIBs observed by SEGUE \citep{Lan15}. DIBs studied by Gaia-ESO are crossed-out, while the ones measured by GALAH are embedded in squares. It is clear that the current surveys are studying only the brightest DIBs within their wavelength intervals. Some of the DIBs, notably the one at 6614 \AA\ are studied by more than one survey, but generally this is not the case.  

Figure \ref{fig_dib_galah} illustrates the most prominent DIBs in the GALAH survey spectra. A number wavelength ranges from a spectrum of TYC 4011-102-1, a hot star with strong interstellar absorptions close to the Galactic plane, are plotted. Individual panels show a dozen DIBs, while the right-most panel plots the interstellar atomic K~I line. The latter in fact reveals the presence of two interstellar clouds with K~I absorption at different radial velocities. For a majority of GALAH targets which lie away from the Galactic plane such complications are rare (but can be detected). 

\section{DIB measurement}

Traditional measurements of DIBs have been limited to hot stars with spectra having only a small number of stellar spectral lines which are thermally or rotationally broadened. Measurement of DIBs in such high S/N spectra is relatively easy, as the DIBs usually lie on a nearly featureless stellar continuum 
\citep[for a discussion of its automation see][]{Puspitarini13}. But observations of DIBs in large stellar spectroscopic surveys draw their strength from a huge number of studied objects. So DIBs should be measured also in spectra of cooler stars, as hot objects are too scarce for the task. As an illustration we note that in a magnitude limited (but otherwise randomly sampled) RAVE survey only 1\%\ of observed stars are hotter than 8100~K and only 1 star in 1000 is hotter than 16000~K. 

Spectra of cool stars are rich in lines with widths that are comparable to those of DIBs. So the latter are almost always superimposed on a rich intrinsic stellar spectrum. Subtraction of the stellar contribution which is needed prior to DIB measurement can be done in two ways. Either we use a theoretical model of stellar atmosphere to compute it or we infer the stellar contribution from other stars' spectra that are very similar but have a different or even negligible presence of the DIB feature. The former approach was used by \citet{Puspitarini15} for Gaia-ESO data, and by \citet{Zasowski15} for the APOGEE set. Unfortunately line lists, values of oscillator strengths, non-LTE effects and the three-dimensional nature of dynamics in stellar atmospheres is still a challenge for theoretical models. Observational apparatus with its residuals of interference fringes, and uncertain level of continuum makes the computation of a realistic stellar spectrum even more difficult. So the studies using this approach understandably focused on sight-lines with strong DIB features, usually these are distant objects close to the Galactic plane. 

An alternative approach, introduced by \citet{Kos13}, exploits a core strength of large spectroscopic surveys, namely that they observe many stars with very similar intrinsic spectra. Such close neighbours are identified by morphological comparison of rest-frame spectra in the wavelength ranges that do not include the studied DIB. So one can assume that these spectra are very similar also within the DIB's wavelength range. The measured difference is then attributed to DIB's contribution. This approach does not depend on physical modelling of stellar atmospheres, even systematic effects of the observational setup tend to cancel out, though a better renormalisation of the star-subtracted spectrum is often needed. Note that a large sample of stellar spectra is needed to find the matching spectra unaffected by the interstellar medium. Nearest neighbour algorithms are applied to the spectra themselves with minimal reliance on stellar parameters, so the results do not require generation of synthetic spectra. In magnitude limited surveys with an otherwise random selection of observed spectra the sample size should be close to a hundred thousand or more for a satisfactory performance of the nearest neighbour algorithm. Note that several surveys listed in Table \ref{tablesurveys} satisfy this requirement. 

\begin{figure}[]
\resizebox{\hsize}{!}{\includegraphics[clip=true,angle=0]{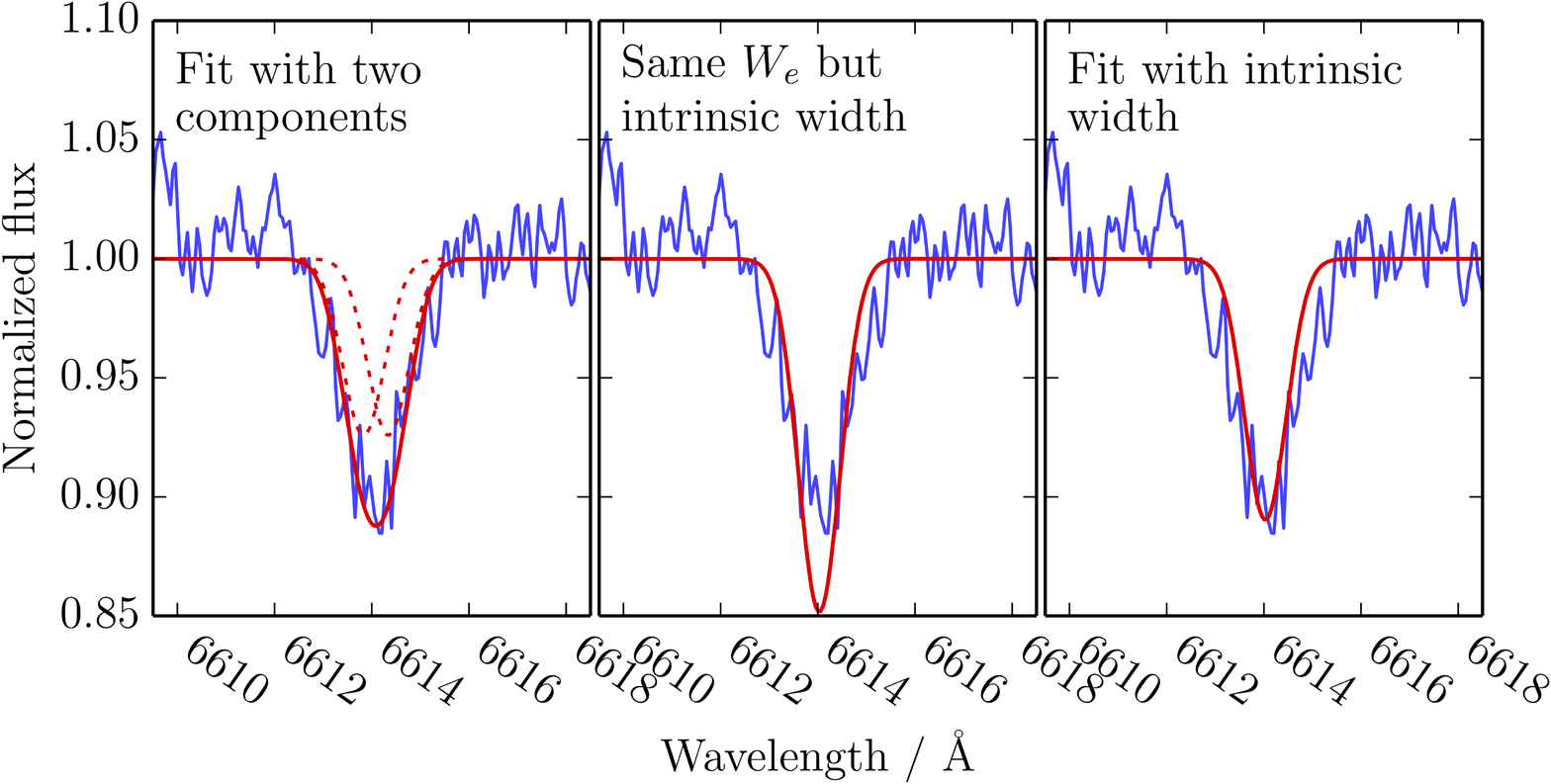}}
\caption{Example of a DIB indicating presence of two absorbing clouds at different radial velocities as observed by the Gaia-ESO survey. The left panel is a fit with two absorbing clouds, the middle one keeps the same equivalent width but uses a single cloud, while right panel is an unconstrained fit using a single cloud.}
\label{fig_twoclouds}
\end{figure}

Diffuse interstellar bands are resolved features, as their name implies. In Section \ref{secDIBs} we mentioned that their profiles can be, in fact, quite complex, due to the presence of partly resolved molecular rotational branches of implied complex molecular carriers. So the spectrum of DIBs obtained after division by the underlying stellar spectrum can be written as
\begin{eqnarray}
F(\lambda) &=& \prod_{i=1}^{D} \prod_{j=1}^{C} \prod_{k=1}^{P} 
  [1 - G(A_{ijk}, \lambda c_{ik}, \sigma_{ik}, v_j) (\lambda)]
\nonumber \\
 & & 
\label{eqDIBprofile}
\end{eqnarray}
where the products go over $D$ DIBs, $C$ interstellar clouds with distinct radial velocities along the line of sight, and $P$ components of the profile of each DIB. $G$ is the adopted shape of the DIB absorption component, e.g.\ a Gaussian with a given amplitude $A$, rest wavelength $\lambda c$, width $\sigma$, and radial velocity $v$. Here we assume that only radial velocity and amplitudes of individual components of a given DIB change from cloud to cloud, while the relative position of components stays fixed which is true if components are partially resolved branches in the electronic transition of a large molecule. 

\begin{figure}[]
\resizebox{\hsize}{!}{\includegraphics[clip=true,angle=0]{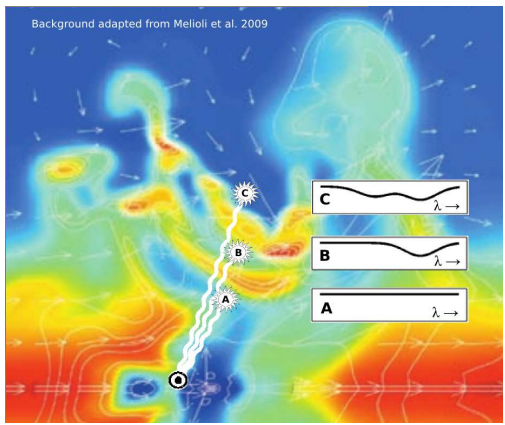}}
\caption{Concept of a multi-dimensional localisation of DIB absorption clouds in the ISM.}
\label{fig_4D}
\end{figure}

Complexity of DIB profiles, as described by eq.\ \ref{eqDIBprofile}, is usually not supported by the data. For noisy observations one usually starts with a simplistic assumption of a single Gaussian and allows for more complicated choice only if a sufficient S/N and resolving power justify it. \citet{Kos13} for example used asymmetric Gaussian profiles, and Figure \ref{fig_twoclouds} shows evidence for two absorbing clouds at different radial velocities. 

DIBs are weak absorptions, so saturation effects which can be very important e.g.\ for interstellar absorption in Sodium doublet \citep{Munari97} are usually negligible. \citet{Zasowski15} present an impressive example of linearity for the DIB at 1.57~$\mu$m even for interstellar extinctions of $A_V \approx 10$~mag. \citet{Puspitarini15} obtain a similar result for optical DIBs observed by the Gaia-ESO survey in a few sight-lines close to the Galactic plane.

\section{Toward a multi-dimensional picture}

\begin{figure*}[]
\resizebox{\hsize}{!}{\includegraphics[clip=true,angle=0]{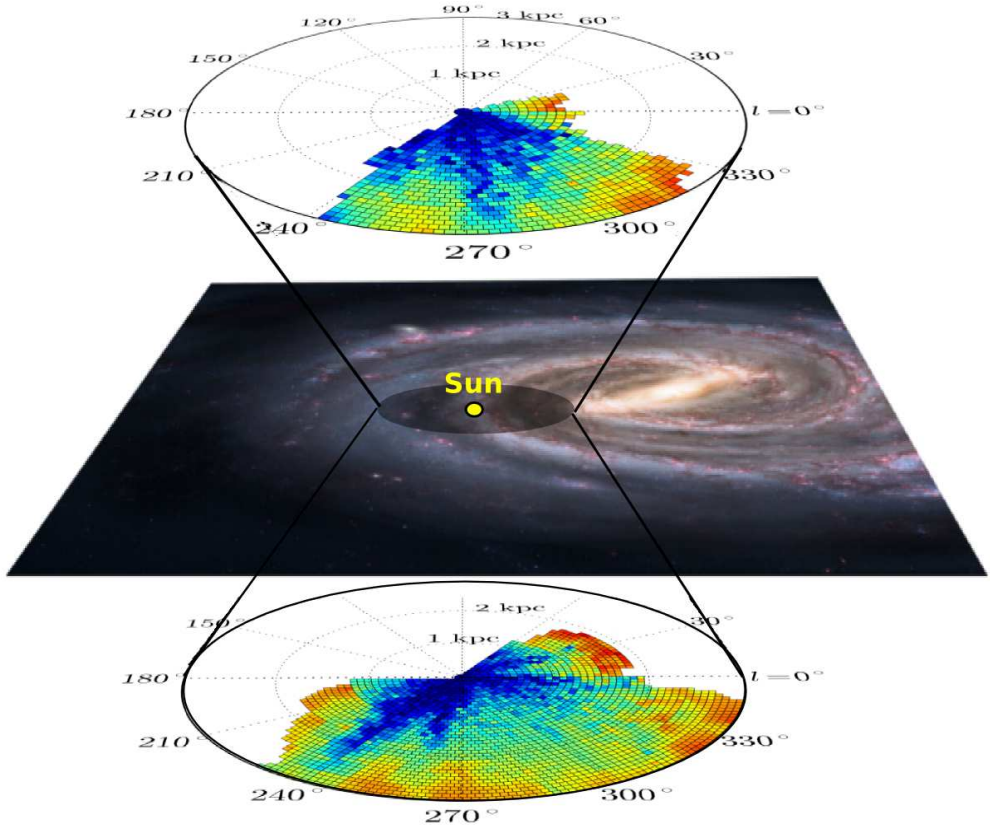}}
\caption{Projected equivalent width of the DIB at 8620~\AA\ for regions on both sides of the Galactic plane, as revealed by the RAVE survey. The color scale is linear with the darkest red tones reaching an equivalent width of $\sim 1.5$~\AA. Adapted from \citet{Kos14}.}
\label{fig_Science}
\end{figure*}

Observation of a single star places the observed DIB absorption somewhere along the line of sight towards the star, but does not fix its exact location. But an ensemble of stars observed in a similar direction and at different distances allows for a spatial localisation of absorbing clouds. In figure \ref{fig_4D} it is clear that while there is no absorption closer than star A we have an absorbing cloud between stars A and B, and yet another one till star C, in this case detected both by an increased equivalent width and by the radial velocity shift of the absorbing cloud. By joining a large number of lines of sight one can finally assemble their distance-resolved clouds into a full 3-dimensional distribution, eventually even measuring radial components of their velocity vector.

Two points should be mentioned when speaking about ISM in multiple-dimensions. First, distances should be known with sufficient accuracy, and clearly this will be a significant contribution of Gaia. At the moment we should resort to spectro-photometric distances, an example are results from a Bayesian approach for the RAVE survey by \citet{Binney14}. Errors on spectrophotometric distances are typically around 20\%. If location of a certain absorbing cloud is revealed by a number of stars their individual distance errors would partially cancel out. So errors in spectrophotometric distances are important but not really a single most important limiting factor in building of a multi-dimensional picture of ISM absorptions. 

The second point is a requirement for a sufficient density of lines of sight. In the GALAH survey there are some 400 optical fibers placed within a field of view of $\pi$ square degrees. So a mean distance between fibers is around 5 arc~minutes, corresponding to 1.6~pc at a distance of 1~kpc. So a combination of accurate astrometric distances from Gaia and a large number of high S/N spectra from surveys like GALAH will allow a 3-dimensional study of ISM at a $\sim 1$~pc resolution. We did not reach this goal yet, both because the distances are not accurate enough, and because in surveys before Gaia-ESO or GALAH the achieved S/N of collected spectra was frequently too low for measurement of DIBs in individual spectra, so that results for many stars at similar distances had to be joined together to improve the reliability of DIB measurement. 

At the moment there are two quasi 3-dimensional maps of the Galactic DIB absorptions. We call them quasi, because they take the distance, the hardest to measure parameter, into account -- but a good distance sampling is compensated by a poor sampling of one of the other dimensions. In the case of \citet{Kos14} this is the Galactic latitude, where the distribution is given only by two independent coeficients and the measured scale height. Two existing maps are covering the southern skies away from the Galactic plane \citep{Kos14} and the northern skies close to the Galactic plane \citep{Zasowski15}. The first one studies the DIB at 8620~\AA, as revealed by the RAVe survey, while the second is an APOGEE study of the DIB at 1.57~$\mu$m. Figure \ref{fig_Science} illustrates the results of the former study. Clearly, absorption increases with distance, as expected, but the picture is not the same for stars above or below the Galactic plane. This shows that DIB absorbers are not completely mixed in the vertical direction, which indicates that their where-abouts are related to relatively recent events in the ISM, e.g.\ supernova explosions not older than a fraction of the vertical oscillation time of $\approx 10$~Myr. The same study also shows that the vertical scale-height of carriers for the DIB at 8620~\AA\ is $118 \pm 5$~pc, which is significantly smaller than the vertical scale-height of dust ($209 \pm 12$~pc). Study of \citet{Zasowski15} finds a very similar vertical scale height of $108 \pm 8$~pc for the DIB at 1.57~$\mu$m. They also construct a spatial map of DIB absorptions projected to the Galactic plane. Their map reaches $\sim 3$-times larger distances than the map of RAVE, but at a lower spatial resolution. On the other hand they were able to map a general velocity variation of the DIB across the sky and confirm that it follows the general differential rotation pattern of the disk. \citet{Lan15} constructed a 2-dimensional map of DIB absorptions off the plane based on SDSS data. All these studies are complementary to each other, as they sample different hemispheres and different distance ranges. 

Construction of 3-dimensional maps of DIB absorptions assumes that the carriers are distributed smoothly in the general ISM. This assumption is generally true, the only exception seem to be particular environments of Herschel 36 (\citealt{Dahlstrom13,Oka13}; but see \citealt{Bernstein15}), and the Red Square Nebula \citep[MWC 922][]{Zasowski15a}. There is also a recent statement of DIB at 6613~\AA\ to be detected in emission in the field \citep{Burton15}.  The argument about DIB absorptions not being related to circum-stellar environments can be turned around by saying that the ISM sampled along a closely spaced set of lines of sight until a given distance should exhibit very similar DIB properties. An example for such a case are stellar clusters, where spectra of all cluster members should include nearly identical DIB absorptions. \citet{Kos15} used the Gaia-ESO data to demonstrate that this is indeed true for most stars considered to be members. A few stars which were considered to be cluster members based on their position in the sky, proper motion, radial velocity, and values of stellar atmosphere parameters, however showed a significantly different strength and/or radial velocity of the DIBs in their spectra, compared to other cluster members. This implies that they are foreground or background stars. So DIBs can present an important additional rejection criterion for cluster membership. On the other hand rich clusters can be used to study granulation of DIB properties in directions perpendicular to the line of sight and on sub-parsec scales. 

\section{Dynamical Interstellar Medium}

Local bubble is a cavity in the ISM with a size of at least 100 pc in the Galactic plane which contains also our Solar system \citep[for a recent review see][]{Lallement14}. Density of neutral hydrogen in the bubble is ~0.05 atoms/cm$^3$, which is about 10 times less than typical for ISM in the Galactic plane. Hot X-ray emitting diffuse gas in the bubble has been recently discussed by \citet{Galeazzi14}, while far-UV radiation has been observed by CHIPSat \citep{Hurwitz05}. A survey of both hemispheres in Na I and Ca II lines has been done by \citet{Welsh10}. The bubble is filled with ionised hydrogen gas at a million degrees embedded in a wall of dense cold gas. It is worth probing such a medium with absorptions in Ca~II, Na~I and K~I atomic absorptions, and also with DIBs which are seen even in relatively harsh environments \citep{vanLoon09}. The latter has been attempted for the northern hemisphere by \citet{Farhang14}. \citet{Berghofer02} provide evidence that the Local bubble must have been created and shaped by multi-supernova explosions. They analysed the trajectories of moving stellar groups in the solar neighbourhood and found that about $\sim 19$ supernovae must have occurred during the past $\sim 14$ million years since bubble's creation. \citet{Fuchs06} show that their implied energy input is sufficient to excavate a bubble of the presently observed size. For a review of recent events connected to the Local bubble see \citet{Lallement15}. 

The ongoing large stellar spectroscopic surveys permit to widen our view beyond the Local bubble. ISM is the place of violent collisions of supernova shells, plus winds from asymptotic giant branch stars and hot-star associations. A typical line of sight of the current surveys which lies close to the Galactic plane penetrates many of these structures, so it is difficult to present an interpretation which reaches beyond the expected increase of the DIB strength with distance and its general correlation with dust extinction. Interpretation of ISM dynamics is even harder, as one can expect that each line of sight samples several head-on collision regions in the Galactic plane. So it is important to note that observations of the on-going GALAH and partly Gaia-ESO surveys are away from the Galactic plane where interactions generally result in a net motion perpendicular to the plane. If any shells of absorbing material are identified we can assume that their motion is perpendicular to shell surfaces and reconstruct a complete velocity vector from its radial velocity component. Such information for ISM is then equivalent to the one collected for stars by Gaia.

This information can be used to study past events in the interstellar medium. They could also identify and characterise Galactic fountains blown away by supernovae in the last million years. Such flows are thought to sustain star formation in the disk by entraining fresh gas from the halo, so they provide a mechanism which explains why star formation in our and other similar galaxies did not stop when gas present in the disk has been used up \citep{BlandHawthorn2009,Fraternali2014}. So dynamical spatial mapping of DIBs observed in the ongoing large stellar spectroscopic surveys is likely to provide an observational calibration of the recent ever more realistic simulations of dynamical evolution of the ISM \citep{Khoperskov14,Girichidis15}.

\begin{acknowledgements}
The authors acknowledge a fruitful collaboration with the other members of the RAVE, Gaia, Gaia-ESO and GALAH surveys. This work has been supported by the  Slovenian Research Agency. 
\end{acknowledgements}

\bibliographystyle{aa}

\end{document}